\documentclass[conference]{IEEEtran}
\makeatletter
\newcommand*{\rom}[1]{\expandafter\@slowromancap\romannumeral #1@}
\makeatother
\usepackage{multirow}
\usepackage[hyphens]{url}
\usepackage{hyperref}
\usepackage[hyphenbreaks]{breakurl}
\usepackage{tabularx}
\usepackage{amsmath,amssymb,amsfonts}
\usepackage{algorithmic}
\usepackage{graphicx}
\usepackage{cite}
\usepackage{textcomp}
\usepackage{xcolor}
\def\BibTeX{{\rm B\kern-.05em{\sc i\kern-.025em b}\kern-.08em
    T\kern-.1667em\lower.7ex\hbox{E}\kern-.125emX}}
\begin{document}

\title{Vulnerability Analysis of Smart Contracts}

\author{\IEEEauthorblockN{Sameep Vani*}
\IEEEauthorblockA{\textit{Ahmedabad University} \\
%\textit{Ahmedabad University}\\
Gujarat, India \\
% https://orcid.org/0000-0002-0392-1847
sameep.v@ahduni.edu.in}
\and
\IEEEauthorblockN{Malav Doshi*}
\IEEEauthorblockA{\textit{Ahmedabad University} \\
%\textit{Ahmedabad University}\\
Gujarat, India \\
% https://orcid.org/0000-0002-0285-2923
malav.d@ahduni.edu.in} \\
\and
\IEEEauthorblockN{Amit A. Nanavati}
\IEEEauthorblockA{\textit{Ahmedabad University} \\
%\textit{Ahmedabad University}\\
Gujarat, India \\
amit.nanavati@ahduni.edu.in}
\and
\IEEEauthorblockN{Ashish Kundu}
\IEEEauthorblockA{\textit{Cisco Research} \\
%\textit{Cisco Research}\\
California, USA \\
ashkundu@cisco.com}
}

\maketitle
\makeatletter{\renewcommand*{\@makefnmark}{}
\footnotetext{* The first two authors have equal contribution}\makeatother}
\begin{abstract}
Blockchain platforms and smart contracts are vulnerable to security breaches. Security breaches of smart contracts have led to huge financial losses in terms of cryptocurrencies and tokens. In this paper, we present a systematic survey of vulnerability analysis of smart contracts. We begin by providing a brief about the major types of attacks and vulnerabilities that are present in smart contracts. Then we discuss existing frameworks, methods and technologies used for vulnerability detection. We summarise our findings in a table which lists each framework and the attacks it protects against.
\end{abstract}


\section{Introduction}
Blockchain technology is useful in securing transactions, assets and computing. It implements a decentralized system in an open peer-to-peer network. It could be considered to be a distributed database. The transactions are governed by various protocols such as proof of work (PoW) and proof of stake (PoS). The transactions are carried through digital currencies (called cryptocurrencies) which are secured by cyrptographic hash functions. This makes blockchain transactions more secure compared to transactions carried through third party, centralized organizations. Although, blockchain and its transactions are more secure than other means of transferring funds, these are not fully secure. In fact, there are numerous attacks that have occurred targeting different aspects of the system.

With the advent of Bitcoin serving as a precursor to popular blockchain networks such as ethereum or polygon, the use of smart contracts (written in Turing-complete languages) is becoming increasingly popular. Furthermore, smart contracts have found applications in many domains including financial industry, public sector and cross industry~\cite{tsankov2018securify}. These smart contracts could become a source of vulnerabilities that an attacker might exploit. Moreover, there can be vulnerabilities in specific third party libraries and packages used. The runtime platform as well as the programming language used can have their own vulnerability issues which can be easily exploited. Not only that, the intangible wallet where the funds are stored, might also contain some security issues resulting in exploitation and illegal transfer of funds. As a result of these vulnerabilities, there have been instances where millions of dollars worth of cryptocurrencies were stolen. One such event was the DAO (Decentralized Autonomous Organization) attack where a reentrancy attack was used to steal \$150M in June 2017. In July 2017, \$30M were stolen from a parity multi-signature wallet and a couple of months later, \$280M were frozen due to a bug in the same wallet.

It can be seen from the above cases that there are various issues such as frozen funds, reentrancy bug etc. that can lead to enormous amounts of money being stolen. Thus, there is a pressing need for early detection of such bugs or vulnerability issues. The main goal of this paper is to briefly discuss the different types of attacks or vulnerabilities that exist in smart contracts and discuss in detail some of the existing frameworks for vulnerability detection, including their architecture, performance, advantages and limitations.

\section{Types of Attacks}

\subsection{DDoS Attack}
According to \cite{IDMDDos}, DDoS (Distributed Denial of Service) attacks are designed to exploit bottlenecks within a system which is accomplished by sending it more traffic than its network can handle. This means that an attacker creates so much traffic (congestion) to a target application that it impedes the traffic flow for legitimate users. Blockchain networks do not have a single point of failure which can be successfully attacked. Even if one of the stakeholder nodes fails, then there are other nodes for backup. This property of decentralized system makes blockchain resistant to DDoS attacks.

However, this does not mean that blockchain networks are fully immune to DDoS attacks. One of the methods discussed in~\cite{halbornDDoS} by which a DDoS attack can successfully affect a blockchain network is by flooding the blockchain with spam transactions. By doing so, an attacker can reduce network's availability for legitimate transactions and carry out a successful DDoS attack.

\subsection{Routing Attacks}
This is one of the most important types of attacks. In this type of attack, blockchain participants typically cannot see the threat and everything appears normal to them. However, behind the scenes, fraudsters have extracted confidential data or currencies.

It is a well known fact that blockchain networks rely on large scale data transfer in real-time. Due to this, the transaction data is transferred via internet service providers (ISPs). Attackers can intercept data as it is being transferred to these ISPs. The attackers hijack BGP (a routing protocol) by advertising its prefix `p'. As a result, all the messages that are passing through the ISP are intercepted by the attacker. The attacker can decide to interfere with the transaction detail by changing the destination wallet address resulting in a theft and stealing cryptocurrencies.

\subsection{Sybil Attacks}
A sybil attack is a type of attack seen in peer-to-peer networks in which a node in the network operates multiple identities actively at the same time and undermines the authority/power in reputation systems \cite{GFGSybil}. There are two types of sybil attacks in blockchain network. In a direct attack, the honest nodes are influenced directly by the sybil node whereas in an indirect attack, the honest nodes are attacked by a {\bf (an honest?)} node which communicates directly with the sybil node. Sybil attacks can result into blockage of users from the network or gives attackers a way to carry out a 51\% attack (see Section~\ref{sec:51} below).

However, many blockchain networks have already mitigated these attacks by using different ``consensus algorithms" to help defend against sybil attacks, such as Proof of Work, Proof of Stake etc.

\subsection{Phishing Attack}
Phishing is a scamming attempt to attain a user's credentials. Attackers send wallet key owners emails designed to look legitimate. These emails ask users for their credentials using fake hyperlinks. This can result in losses for the user and the blockchain network.

Attackers can steal all the cryptocurrencies from a user's wallet and can initiate fraudulent transactions leading the user to pay for the damage incurred. These types of attacks are usually detected by spam detection algorithms which run on each mail delivered to the user. Many email services like Gmail and Yahoo mail already use these spam detection algorithms and hence the risk of these attacks is reduced to a considerable extent.

\subsection{51\% or majority attacks}
\label{sec:51}
In modern blockchain networks, various consensus algorithms are used to arrive at a common state for the blockchain. This means that without the consensus between miners, a block of transaction will not be added to the main blockchain. Thus, mining requires a vast amount of computing power including large hashrates and network resources, especially for large-scale public blockchains.

However, if a miner or a group of miners could rally enough resources, they could attain more than 50\% of a blockchain network's mining power. Having more than 50\% of the power means having control over the ledger and the ability to manipulate it~\cite{8563187}. Also note that private blockchains are not vulnerable to 51\% attacks because they have specially elected miners called stakeholders who arrive at a consensus. There have been various attempts for 51\% attacks. One of them was attempted on bitcoin network in August 2021 where 5\% of Bitcoin SV (BSV) were stolen~\cite{economicTimes51}.

\subsection{Reentrancy attack}
A reentrancy attack is one of the most destructive attacks in solidity smart contract. A reentrancy attack occurs when a function makes an external call to another untrusted smart contract. As a result, the untrusted contract makes a recursive call back to the original function in an attempt to drain funds.

When the contract fails to update its state before sending funds, the attacker can continuously call the withdraw function to drain the contract's funds. One of the most notable reentrancy attack was the DAO attack on Ethereum Classic network~\cite{mediumDaoStory}. The attack stole \$150M. 

\subsection{Double spending attack}
A double spending attack as described in~\cite{rosenfeld2014analysis} as a successful attempt to first convince a merchant that a transaction has been confirmed and then convince the entire network to accept some other transaction. As a result, the merchant would be left with neither the product nor the coins and the attacker gets to keep both.

In May and June 2018, a double spending attack was successfully carried out which lead to the stealing of \$18.6M worth of bitcoins. Many efforts have been made to resolve the double spending attacks. One such model is described in~\cite{begum2020blockchain}.

\subsection{Smart contract overflow and underflow}
According to~\cite{CryptoAdventure}, overflow and underflow attacks are often classified as integer overflows. These errors allow a program to revert to the start or to re-calibrate. An overflow error attack on a smart contract occurs when more value is provided than the maximum value resulting into a circle back to zero. On the other hand, underflow error attack occurs when you go below the minimum amount triggering the system to bring you right back up to the maximum value instead of reverting to zero. Between the two, underflow attacks are simpler to carry out.

These errors can be disastrous. If a program lacks the feature that checks for underflow and overflow, an attacker can get more tokens than they own. They can also get a maxed-out balance, which is essentially stealing. Furthermore, these attacks can cause the whole system to break because the amount of tokens being maxed out is not the same as the tokens in the system.

\subsection{Short address attack}
This type of attack is due to an infamous vulnerability issue in Ethereum Virtual Machine (EVM). Short address attack occurs when a contract receives less data than it was expecting~\cite{mediumShortAddress}. It occurs when there is inadequate or no validation being performed on the length of transaction payload. This tricks smart contracts into performing malicious transactions.

\subsection{Transaction Ordering Dependence}
A transaction-ordering attack is a race condition attack. The attack changes the price during the processing of a transaction T because either the contract owner, miner or another user has sent another transaction, modifying the price, before T is complete.

As a result, two transactions can be sent to the mempool along with the gas sent with the transaction. This gas determines which transaction will be mined first. An attacker, thus, will choose the order in which the transactions are mined. This creates a problem in Smart Contracts that rely on the state of storage variables to remain at certain value according to the order of transactions. 

\subsection{Exception Handling}
This vulnerability arises when the contract has some mishandled exceptions. One of the instance of handled exceptions is the ``King of Ether" contract, whose creator publicly asked users to not send ether to this contract since the return value of \textbf{call} was not checked but returned successfully.

This violation can be observed if there is a different behavior depending on the result of the call.

\subsection{Frozen Funds Attack}
It is an attack where an ethereum contract is removed from a blockchain using a particular \textbf{kill instruction}. In this case, the attacker removes the library associated with a wallet. This leads to the freezing of funds and making them inaccessible.

This attack is often caused by the dependency of wallet on a library. In other words, there is no guarantee that ether after being deposited can be transferred out of a particular contract since this transaction is dependent on the library. In an attack in late 2017, a wallet was removed since it had frozen funds worth 280 million dollars \cite{tsankov2018securify}.
According to \cite{tsankov2018securify}, there are two behaviors to look for: \emph{(i)} users can deposit their ether and \emph{(ii)} the contract has no call instruction for transferring them when there is a non-zero ether amount.

\subsection{Inifinite Loop Attack}
This vulnerability arises when a program gets stuck in a loop and can never break out of it \cite{mediumInfiniteLoop}. This vulnerability, in itself, is not dangerous but it has some serious consequences. It can cause the program to crash leading the deducted cryptocurrency to get lost. It can also consume a lot of computing power and resources of the system causing a Denial of Service attack. This vulnerability can cause problems for the user if attackers decide to exploit it.

\subsection{Timestamp Dependence}
According to \cite{githubTimestampDependence}, an attacker can manipulate the timestamp within 15 seconds of block validation. As a result, the attacker can change and set the value of timestamp that would increase their odds of benefitting from the function. This is only possible if a smart contract uses block.timestamp or block.number. Furthermore, the attacker can trigger time-dependent events as in a decentralized system, precise timestamp is very difficult to provide \cite{swcTimestamp}.

\subsection{Callstack Depth}
Callstack depth is a vulnerability that can make an external call fail because the call exceeds the maximum call stack of 1024 \cite{mediumCallStack}. Since the stack is full, the call will fail and if exception is not handled properly, then it can lead to an output that favours the attacker.

\subsection{Unchecked Send}
This vulnerability is one of the consequences of callstack depth attack mentioned above. To send Ether to some hashed address, ``send" method is usually invoked. However, this send() method can fail due to two reasons: \emph{1.} The address given is for a contract and not user account. This contract can generate an exception resulting in loss of transferred Ether. \emph{2.} Due to limited resources, callstack depth vulnerability might take place which can result into loss of Ether \cite{uncheckedSend}. Hence, it is generally advisable to check for such issues before sending cryptocurrency.

\section{Approaches}
\label{section3}
In this section, we aim to curate multiple approaches that exist in detecting above mentioned security attacks or vulnerabilities. It becomes immensely important to have knowledge about existing frameworks and compare various metrics with each other in order to determine which framework is best suited for detecting which type of vulnerability or attack.

\subsection{Securify}
Securify \cite{tsankov2018securify} is an open source vulnerability detection framework for ethereum smart contracts supported by Ethereum Foundation and Chain Security. It takes different instances of fraud into account and it has created a customizable compliance and violation patterns. It has achieved successful automation of most of the processes. Input to the securify program is EVM byte code and pre-defined security patterns that are generated by securify based on a lot of fraud instances.

Securify detects vulnerabilities in a three-step process. In the first step, the EVM bytecode is converted to decompiled code. Following this, in step 2, semantic facts are extracted from this code using rule-language datalog (used in stratified format). In the final step, securify checks for compliance and violation patterns and identifies them giving results of the existing vulnerabilities in the smart contract. There are many patterns already detected and added to securify database of patterns. Some of these include ether liquidity (LQ), transaction ordering dependencies, reentrancy, restricted writes (RW), restricted transfer (RT), handled exceptions (HE) and validated arguments (VA).

Securify has been evaluated on solidity dataset obtained from ethereum website. On running securify, it has been observed that the highest level of false warning is seen in RT and TA, followed by TR and LQ. We can conclude that securify does well in identifying potential vulnerabilities provided that it has pre-defined violation patterns for each and every vulnerability. However, there are some shortcomings too. Firstly, it assumes that all instructions are reachable irrespective of whether they are actually reachable or not. Secondly, it fails to capture contract-specific requirement. It lacks a refined warning system. Lastly, properties or patterns can also be exploited by the attackers.

\subsection{Vandal: A scalable security analysis framework for smart contracts}
\cite{brent2018vandal} presents a tool for vulnerability detection of smart contracts. It detects five different types of vulnerabilities namely unsecured balance, reentrancy, unchecked send, destroyable contract and use of ORIGIN. It expresses security analysis in a declarative fashion (logic specification). It uses Souffl{\'e} declarative language \cite{jordan2016souffle}. Vandal uses Datalog as a domain specific language to connect program semantics of security vulnerabilities and implementation of corresponding analysis. Finally, these specifications are executed by a datalog engine and a list of detected security vulnerabilities and their locations in the bytecode are produced as output. As a result, vandal has been divided of two parts. First, an analysis pipeline that translates low level code to logical relations (which are used to find data and control flow dependencies in the bytecode). Second, a set of logic specifications for security analysis is used.  In order to implement this, vandal uses Souffl{\'e} \cite{jordan2016souffle} as a datalog engine. Furthermore, vandal proposes a new decompilation technique by applying symbolic execution on basic blocks.

The vandal framework subdivided into several stages. The first one is the \emph{scraper} which helps in fetching the ethereum bytecode from the blockchain. Next is a \emph{disassembler} which is useful to convert bytecode into mnemonics. A decompiler is the next stage that translates the stack-based bytecode to a register transfer language. Finally, the extractor converts intermediate representation of the register-transger language into logical relations. These are then used to indentify semantics of the smart contracts. Moreover, Souffl{\'e} generates executable programs from security analysis for detecting vulnerabilities.

For evaluation purposes, runtime is considered as a performance metric. It has been shown that, on an average, vandal reports vulnerabilities in 4.15 seconds whereas other tools such as Oyente, Ethir, Mythril and Rattle take 13.68 seconds, 11.99 seconds, 11.10 seconds and 4.47 seconds respectively.

\subsection{BASNEA: Threat Hunting for Ethereum Smart Contract Based on Backtrackless Aligned-Spatial Network Entity Alignment}
\cite{du2022basnea} proposes a backtracking algorithm to identify attacks, suspicious and benign behaviours. It does so by comparing the attack provenance graphs constructed by the Ethereum threat intelligence with transaction provenance graphs generated from ethereum sync node.

BASNEA is a hybrid approach that uses threat hunting and Graph Convilutional Neural Networks. Threat hunting is the process of actively and repeatedly searching the network for the sake of detecting and isolating advanced threats that evade existing security solutions\cite{du2022basnea}. The notable contribution of this tool is that it considers both attacked Dapp and attacker's perspective. The whole process employed by BASNEA is divided into 5 stages.

The first stage is preprocessing the data. This also includes fetching the data from various sources. This is accomplished through a series of extractors. The first one is called Ehtereum Transaction Extractor which uses API interface to query transactions. Second is attack behaviour extractor and validator. Its job is to collect threat intelligence information from CVEs, Dapp vendors announcements, security company's blog and researcher blogs. Lastly, the collected information is used in second stage.

In the second stage, transaction provenance graph (a hypergraph) is constructed. The nodes of the graph includes all accounts in the transactions and smart contracts. Edges are created between two nodes that have some relation with each other such as ether transfer, smart contract invocation or suicide etc. Once, constructed, the third stage of the tool starts which includes running BASNEA algorithm. Lastly, BASNEA finds the attack objects and attackers described in the threat intelligence. During this last stage, the paper discusses the use of principle component analysis to reduce the dimensions and k-medoids clustering algorithm to cluster suspicious behaviours into difference attack methods.

For the evaluation, the dataset is obtained from ethereum transactions, CVEs and dapp vendors. Paper compares BASNEA with various methods used for vulnerability such as RNN, GCN etc. It is observed that the overall F1-score achieved by BASNEA is 0.93. Apart from this, BASNEA has been analysed on number of attacks reported by BASNEA within 1 day, 3 days, 7 days and 30 days. Based on the analyses, it is observed that 5\% of the attacks were concentrated in 3 days after the application went online. Also, Finally, analysis is done from attacked Dapp's perspective and attacker's perspective for 4-5 different vulnerabilities.

\subsection{ESCORT: Ethereum Smart Contracts Vulnerability Detection}
ESCORT \cite{lutz2021escort} or Ethereum Smart Contract vulnerability Detection is an efficient and scalable multi-vulnerability detection framework for smart contracts. It employs a multi-output architecture where the feature extractor learns the program semantics and each branch of the DNN captures the semantics of a specific vulnerability class. It uses bytecodes of publicly available smart contracts which reduces the overhead of understanding semantics of smart contracts written in different languages which is a problem with source code. To label each vulnerability of smart contract, it uses existing frameworks like Securify, Mythril etc. and uses the label given by that framework which has the highest F1 score. Deep Learning algorithm uses two methods namely Text Representation and RNN. In the former one, the text modality is transformed into numerical vectors for usage in ML algorithms. The latter one is used to construct a direct computational graph along the temporal sequence.

Along with that, ESCORT supports lightweight transfer learning on new vulnerability types and hence is extensible and generalizable.  The architecture is divided into three phases which are training phase, transfer learning phase and deployment phase.

The training phase enables supervised learning for vulnerability detection. The model initially constructs the smart contracts bytecode dataset with corresponding labels. Then, it specifies the system parameters, including the vulnerabilities of interests and available hardware resources for ESCORT's multi-output DNN design. Finally, the devised model is trained on the collected contract data with their corresponding labels which results into a converged DNN detector.

The next phase is the transfer learning phase. It is able to generalize to new vulnerabilities through this phase. It creates a new branch of the pre-trained DNN model and passes the output of the feature extractor to all the branches. Finally, it uses knowledge from other branches to adapt and detect new types of vulnerabilities.

The last phase is the deployment phase. After training and transfer learning phase completes, ESCORT returns a trained DNN classifier that can detect whether an unknown smart contract has any of the learned vulnerability types.

Evaluation of ESCORT suggests that it does a really good job (based on F1-score) in detecting several already existing vulnerabilities. It has successfully achieved 0.95 as the F1-score on test dataset. Furthermore, it is able to generalize pretty well and does a great job in detecting new vulnerabilities. From the results as explained in \cite{lutz2021escort} we can conclude that even though ESCORT's FPR (False Positive Rates) and F1-score are a bit degraded when compared to a model that is trained from scratch, the time taken by the model to train from scratch is well compensated here.

Overall, ESCORT is the first framework to use Deep Learning models. It uses concepts of NLP and RNN to train the model using labeled bytecode. It solves the issue of the existing framework's flexibility to adapt to new types of vulnerabilities using transfer learning and branched architecture. It addresses issues of scalability of existing frameworks by employing multi-output architecture where the feature extractor learns the program semantics and each branch of DNN captures the semantics of a specific vulnerability.

\subsection{Sequence Learning Approach}
\cite{tann2018towards} is yet another tool based on machine learning. It is based on LSTM (Long Short Term Memory) which allows it be able to detect new attack trends quickly. As a result, it is able to generalize to newer vulnerabilities quickly compared to conventional methods. It has been mainly compared with symbolic checking tools such as MAIAN \cite{nikolic2018finding} and Oyente \cite{luu2016making}. The main objective of the paper is to build a vulnerability detection platform using LSTM that required constant analysis time as smart contracts grow in complexity. This till can easily scale to process a large number of smart contracts. Furthermore, it is the first framework to apply LSTM model for detecting smart contract security threats at an \emph{opcode level}. The usage of opcodes is supported by evidences of its use for malware detection in windows and android.

The classification process begins by generating labels required for supervised machine learning. The contracts are processed by passing bytecodes through MAIAN \cite{nikolic2018finding}. In the same process, opcodes are also retrieved. The opcodes serve as an input to the sequence learning model. Furthermore, these opcodes are modelled with code vectors. Code vectors capture potential relationships in sequences and these sequences are, in turn, learnt by LSTM.

The dataset is downloaded from Google BigQuery and is parsed using EVM instruction list \cite{wood2014ethereum}. As mentioned above, this dataset was passed through MAIAN \cite{nikolic2018finding} to obtain classes. However, this dataset was highly imbalanced. In order to balance these, all the vulnerable contracts were grouped together in one class. Hence, the problem was reduced from multi-classification to binary classification. Moreover, the dataset was resampled to achieve a balanced distribution. Technique called SMOTE (Synthetic Minority Oversampling Technique) was used where they oversampled minority class (vulnerable) and undersampled majority class (non-vulnerable).

Experiments demonstrates that LSTM achieved a test accuracy of 99.57\% and an onverall F1 score of 0.86. Inspite of great accuracy and F1-score, it does have some limitations. The model assumes that the sequence of opcode features can generalize its content and its vulnerabilities. However, this assumption is violated multiple times. It does not support the checking of data and control flow properties\cite{tann2018towards}. In such situations where the assumption does not hold true, LSTM will fail to perform well against vulnerable smart contracts. Second limitation of this approach is that it is dependent on MAIAN for initial vulnerability classes. However, these classes might not be enough to describe each vulnerability that exists. There are more vulnerabilities that need attention and hence there is a pressing need to improve and accommodate all these vulnerabilities.

\subsection{Smart Contract Vulnerability Detection Using Graph Neural Network}
As suggested by the name, this approach uses graph neural network for detection of vulnerabilities in smart contracts \cite{zhuang2020smart}. It also has an open source implementation. The most important aim of this paper is to resolve two of the major concerns of existing methods. Firstly, existing methods heavily rely on several expert-defined hard rules to detect smart contract vulnerabilities. However, expert rules are error-prone and some complex patterns are non-trivial to be covered. Secondly, since the rules are contributed by a few ‘centralized’ experts who develop the detection tools, their scalability is inherently limited. To address these issues it characterizes the source code of a smart contract as a contract graph according to data and control dependencies between program statements.

The method in paper is divided into three phases. First is a graph generation phase which extracts the control flow and data flow semantics from the source code and explicitly models the fallback mechanism. Second is a graph normalization phase inspired by k-partite graph. Third is a novel message propagation networks for vulnerability modeling and detection.

In the graph generation phase, the model formulates a smart contract function into a contract graph and assign distinct roles to different program elements. Further, it constructs edges by taking their temporal order into consideration. Moreover, the nodes in the graph are categorized into three categories. First, major nodes which symbolize the invocations to customized or built-in functions that are important for detecting specific vulnerability. Second, secondary nodes which are used to model critical variables like user balance or bonus flags. Lastly, fallback node are constructed to stimulate the fallback function of an attack contract which can interact with the function under test.

In the graph normalization phase, the model removes each secondary node but passes the features of it to its nearest major node. If a secondary node has multiple nearest major nodes, the features are passed to all of them. The fallback node is also removed similar to the secondary node. 

In the last phase i.e the message propagation network phase for detecting vulnerabilities, messages are passed along the edges, one edge per time step. At each step, the hidden state for each node is updated. Finally, at the end step, the end node updates its hidden state by aggregating information from the incoming message and its previous state. Finally, after successfully traversing all the edges in the normalized contract graph,  the model computes a label for the graph by reading out the final hidden states of all the nodes. During training, networks are fed with a large number of normalized contract graphs constructed from smart contract functions, together with their ground truth labels. The, the trained models are employed to predict the vulnerabilities in unseen normalized graph.

The evaluation done on this approach suggest that GNN-based approach does a pretty descent job on detecting multiple vulnerabilities such as reentrancy, timestamp dependence and Infinite loop with F1-score of 0.7811, 0.7919 and 0.7410 respectively. It has also been compared with various other tools such as oyente, mythril and security and it is observed that it outperforms these.

Even though GNN-based method supports multi-class detection, it has the following limitation as described by \cite{lutz2021escort}. Since it uses source code to generate graph, it has restricted applicability because source code are hard to obtain from public blockchain.

\cite{liu2021combining} extends the above discussed GNN based method. It combines the Graph Neural Network with expert defined patterns. \cite{liu2021combining} paper explicitly models the key variables in the data flow. It extends message protocol delivering higher performance. This leads to consistent better performance than the previous work and establishes a new state of the art metrics and performance. The process for vulnerability detection as followed by this paper is \emph{1.} For security patterns that are extracted from the source code, a feed forward neural network is used that generates the pattern features. \emph{2.} In order to extract the graph features from the contract graph formed from the graph generation and normalization phase, a temporal message propagation network is used. \emph{3.} CGE (Combining Graph feature and Expert patterns) network combines graph feature and pattern feature which is the input into the FC and sigmoid layers to output the vulnerability detection results. Based on the results provided in \cite{liu2021combining}, CGE outperforms the previous GNN-based deep learning. Its F1-score is 0.86, 0.87 and 0.82 for reentrancy, timestamp dependence and infinite loop respectively. Its limitations is that the framework uses source code for vulnerability detection which is not easily publicly available. Furthermore, it uses an architecture that could only detect a limited number of vulnerabilities. Hence, there is a need to extend this architecture to generalize and detect more number of vulnerabilities.

\subsection{Ethir: A Framework for High-Level Analysis of Ethereum Bytecode}
\cite{albert2018ethir} presents another framework that analyzes bytecode extracted from ethereum virtual machine. It modifies Oyente's rules that generates Control Flow Graphs (CFGs) \cite{luu2016making}. Hence, it is a rule-based framework for ethereum bytecode. The difference between Ethir and Oyente is that CFG generated by Oyente only stores the last value of jump address when non-unique addresses are given as input whereas Ethir includes all possible jump addresses so that the whole CFG is recontructed. Furthermore, Ethir uses guarded rules representing the continuation when a particular condition from the given set of rule holds true and when it does not. Note that the guarded rules are only used when a conditional jump instruction is encountered. Its limitation includes that Ethir has not been formally evaluated. Hence, the efficiency and effectiveness of Ethir can not be determined as of now.

\subsection{Gasper: Under-Optmized Smart Contracts Devour Your Money}
\cite{chen2017under} introduced another tool called GASPER that automatically locates gas costly patters by analysis bytecodes of smart contracts. Gasper aims at detecting 7 gas costly patterns and clusters them into 2 categories. However, it is currently able to detect 3 gas costly patterns namely Dead Code, Opaque predicate and Expensive operations in a loop \cite{chen2017under}. It uses symbolic execution to cover all code blocks. Gasper first disassembles the bytecode using disasm which is provided by Ethereum. Then, it constructs Control Flow Graph (CFG). Finally, after generation of CFG, the symbolic execution starts and traverses the CFG. It is successfully able to identify 93.5\%, 90.1\% and 80\% \cite{chen2017under} of smart contracts that suffer from these 3 patterns respectively. 

\subsection{Graph Embedding Based Bytecode Matching}
\cite{huang2021hunting} introduces a completely new method of detecting vulnerable smart contracts. It uses an unsupervised graph embedding algorithm to encode the code graphs into vectors. More specifically, it uses graph2vec algorithm \cite{narayanan2017graph2vec} in order to capture structural information. In order to generate Contract Flow Graph (CFG), it uses bytecode of smart contracts. It also implements bytecode-oriented normalization and slicing techniques and a comparison between vulnerable patterns and patterns obtained from CFG are compared.

The idea behind such an approach is to extract patterns from a given smart contract and compare the unknown patterns with vulnerable patterns to see if they match. Once the bytecode is obtained, it is preprocessed to construct CFG based on slice criteria. Later on, the sliced graph is normalized and a finally a graph embedding network is used to vectorize the CFG. Lastly, the obtained vectors are compared with vectors obtained from vulnerable slices and a similarity number is calculated by measuring pair wise similarities. Based on this, potentially vulnerable smart contracts are reported. Note that slicing removes unwanted or unrequired patterns based on certain criterias. Normalization ensures that there are no discrepancies once the bytecode has been sliced. Discrepancies can arise due to different compilers or language used. Two types of normalization are carries namely Data normalization and Instruction normalization \cite{huang2021hunting}. 

After the graph embedding is completed, a corresponding vector is generated for every slice. For a given smart contract and a vector generated previously, a number corresponding to similarity is obtained based on certain mathematical foundation \cite{huang2021hunting}. The larger the values of the number, the higher the similarity between two slices. Finally, results are ranked based on similarities and threshold value.

The evaluation is done on three datasets extracted from various sources like open-source, cve vulnerable datasets etc. Through experiments and tesing, it was observed that average time for one targeted smart contract is 0.47 seconds indicating a high performance rate. False positive rate of the tool is 5.93\% which is very low when compared to Oyente and Mythril.

\subsection{ContractWard: Automated vulnerability detection models for ethereum smart contracts}
ContractWard \cite{wang2020contractward} is yet another tool that utilizes machine learning techniques to detect vulnerabilities in smart contracts. The major aim is to improve the efficiency of vulnerabilities detection in smart contracts ensuring the accuracy of detection. It is a tool that uses multiple disciplines of many fields including machine learning for detection, NLP, blockchain etc. The proposes system provides large-scale and automated vulnerability detection on Ethereum smart contract with machine learning algorithms. Contract ward is sufficient in detecting six vulnerabilities namely Integer Overflow and Underflow, TOD, callstack Depth attack, reentrancy and timestamp dependence.

ContractWard \cite{wang2020contractward} is built with six steps. First, it collects a big number of verified smart contracts. Second, the opcodes are obtained from source codes and are simplified. Third, bigram features and labels are added using existing symbolic checking tool called Oyente \cite{luu2016making}. Fourth, it classifies the dataset by employing One Vs Rest (OvR) algorithms for multi-label classification. Fifth, sampling algorithms are used to balance the dataset before training. Finally, detection models are built on the balanced training sets.

For the first step, approximately 49000+ smart contracts were obtained from official Ethereum website. Note that the dataset collected contains smart contracts that have been verified before September 2018. The class distribution for integer overflow and underflow are balanced whereas for others this is not the case. For labelling the data, Oyente \cite{liu2021combining} is used to produce six labels for each contract. Then manual intervention is done to ensure the correctness of the labels. Note that the labels are independent from each other in each type of vulnerability. Next, bigram algorithm is used for extracting features \cite{cavnar1994n}. Through a sliding window of binary-byte size, opcodes are segmented into massive bigrams \cite{cavnar1994n}. For dimensionality reduction, opcodes are simplified by dislodging the operands and classifying functionally similar opcodes into one category. For the final step in data preprocessing, sampling algorithms like SMOTE or SMOTETomek is used. These algorithms remove the imbalance from the dataset.

Having the dataset ready with labelling completed, dimensions reduced and classes balanced, the next step is traning classification algorithms based on One Vs Rest (OvR) approach. 5 classification algorithms have been used namel eXtreme Gradient Boosting (XGBoost), Adaptive Boosting (AdaBoost), Random Forest (RF), Support Vector Machines (SVM) and k-Nearest Neighbor (KNN). Hyper-parameters have been tuned for better results of all the models. One thing to note is that during training, n-fold cross-validation has not been considered. Finally, based on the performance of all the models on the test dataset, XGBoost classifier on a balanced training sets with SMOTETomek is used in ContractWard.

Through XGBoost, highest F1-score of 0.99 was achieved for Integer overflow and underflow. F1-score for TOD, callstack depth, timestamp dependence and reentrancy are 0.96, 0.97, 0.95 and 0.95 respectively. Overall micro-f1 and macro-f1 scores are 0.9848 and 0.9641. These values, by far, indicate that ContractWard performs the best in detecting these six types of vulnerabilities. However, there are two limitations of the above tool. First, it is dependent on Oyente \cite{luu2016making} for labelling the dataset. As a result, there is a need for manual intervention to check the correctness of labelled dataset. This leads to more time consuming process. Second, ContractWard would not be able to generalize to more number of vulnerabilities because it has been trained on specific six types of vulnerabilities.

\subsection{SmartShield: Automatic smart contract protection made easy}
\cite{zhang2020smartshield} discusses a novel approach, not only to detect, but also to rectify the vulnerable smart contracts. It relies on bytecode based semantic extraction and focuses on fixing three major types of vulnerabilities namely reentrancy, arithmetic overflows and underflows and missing checks for failing external calls. The main aim of the tool is to address issues of scalability, efficiency and gas optimization. The paper \cite{zhang2020smartshield} claims that it is able to rectify majority of the smart contracts with minimal increase in gas fees. Overall, smartshield takes a smart contract as input and generates rectified smart contract free from the above mentioned vulnerabilities. Smartshield contains two phases: \emph{1.} Semantic extraction phase where it extracts semantics of each bytecode by analyzing abstract syntax tree (AST) and \emph{2.} Contract rectification phase where it fixes the control and data flows with the help of semantic information.

For the first phase i.e Semantic extraction phase, smartshield analyzes abstract syntax tree (AST) combined with raw EVM bytecode. Using AST, it first generates control flow graph (CFG) which in turn is used to find data flow dependencies. However, these raw control and data flows can not be used directly. For control flow, it follows a three step procedure.
First, it terminates the execution of smartshield if instruction such as STOP, RETURN etc is found. Second, it transfers the execution to the target address if instruction such as JUMP, JUMPI etc is encountered. Finally, it executes the bytecode normally if none of the above instructions are encountered. As for the data flow, it emulates stack and memory separately. For the second phase i.e Contract rectification phase, it employs two procedures. First is control flow transformation where instructions that change state are moved in front of call instructions. Second is dataguard instruction where instructions that are manually created are inserted to check and validate arithmetic overflows and underflows. Furthermore, minimal number of dataguards are inserted in order to optimize for gas fees. Finally, after completion of both these phases, a rectification report is generated which includes bytecode relocation and bytecode validation instances.

Smartshield has been evaluated on over 50,000 smart contracts. It has been observed that the gas fee for 94.4\% of the smart contracts increased only by 0.1\%. Only a small number of smart contracts have their gas fee increased by a significant amount. Furthermore, manual validation was also done to gain an in-depth insights on the rectification results.

\subsection{Color inspired inspection of potential attacks}
This framework uses convolution neural networks (CNN) in order to detect potential vulnerabilities in smart contracts. The main aim of the paper \cite{huang2018hunting} is to translate the bytecode of solidity into RGB color code. Then, the transformed images are encoded into fixed-size images. The encoded image is fed into CNN (Convolutional Neural Network) for automatic feature extraction and learning, detecting compiler bugs of Ethereum Smart Contract. Moreover, the paper discusses the use of transfer learning for classifying multi-label classification results.

There are several steps listed that the model performs.
\begin{itemize}
\item Crawling the bytecode of smart contract from etherscan by pyspider.
\item Transforming the bytecode of smart contract into RGB color code and transforming them to a fixed sized encoded image.
\item After that, the encoded image is fed to convolutional neural network for automatic feature extraction and learning.
\item Finally, once the model has been trained and validated, the framework deploys it on the backend service.
\item After the bytecode of smart contract are all identified, the scanned results will be provided to the users through UI and public RESTful API.
\end{itemize}

\cite{huang2018hunting} also provides a comparison between accuracies of different pre-trained CNN based models used like AlexNet, GoogleNet etc and that too with varying number of epochs. Based on testing these various models, Inerption-v3 performs the best with almost 97\% accuracy. However, there are certain limitations to this approach. The limitations include its low performance on multi-label classification and its generalizability is not evaluated and discussed.

\subsection{Smart Contract Vulnerability Detection model based on Multi-task learning}
\cite{huang2022smart} developed a smart contract vulnerability detection model based on multi-task learning combining multiple tasks to complete multi-label classification. The main aim of the paper is to address issues of scalability, multi-label classification, non-dependence on source code and generalizability to newer vulnerability types. As a result, the paper introduces multi-task learning into security detection. The model consists of two parts: \emph{1.} The bottom sharing layer using neural network to learn semantic information of input contracts and extract feature vectors and \emph{2.} the specific task layer using CNN to establish a classification network for each task branch.

The dataset of smart contract and related meta data is collected from XBlock platform. This dataset provides the source code along with other additional information such as timestamp, creator etc. Later the data is cleaned to remove irrelevant, vacant and not useful fields. In the next step of preprocessing, the opcode is obtained from bytecode which in turn is obtained from the source code. In order to obtain opcode from bytecode and bytecode from source code, Ethereum Yellow Paper \cite{wood2014ethereum} was used. The imbalance in data is removed by resampling this dataset i.e undersampling contracts without vulnerabilities and oversampling contracts with vulnerabilities.

For training, the model is divided into two parts model namely bottom sharing layer and top specific task layer. The former one is used to achieve knowledge sharing among all tasks in the multi-task learning model. This sharing layer network is based on the attention mechanism to enable the model to learn the features of smart contract sequences. This layer uses concepts of NLP like work embeddings to understand the semantics of the smart contract. Word embedding converts input opcode into a vector to map opcode sequence. After this, positional embedding is done. It adds the position information to the word coding helping to form a new representation for better expressing the distance between the words. The output of word and positional embedding is provided as input to multi-task layer for learning and extracting features. This layer is subdivided into many self-attention layers. Finally, the output of these self-attention layers (or multi-attention layer) is provided an input to a feed-forward neural network. Its job is output the common set of features for all the task making the training easier. Thus, this final layer outputs the results of the feature-sharing layer. The second layer called specific task layer is mainly used for vulnerability detection. There are two sub-tasks that need to be performed namely detection and recognition. For detection (which is essentially a binary classification problem), CNN is used to build binary classifier. Input to the convolution layer is the feature vector obtained from the shared layer. On the contrary, the recognition task is a multi-label classification problem. It also uses CNN to build the task network but more number of neurons are added in the last fully connected layer and softmax activation function is applied to calculate the probability of each output.

For evaluation purposes, four experiments were conducted. In the first experiment, multi-task learning tested different parameters and carried a vulnerability detection experiment. In the second one, model comparison with existing vulnerability was made. In the third one, model was compared with ML-based vulnerability detection tools and lastly the fourth experiment was an ablation experiment. From the first experiment, it was concluded that the batch size of 64 was best in order to find a balance between training time, accuracy, F1-score etc. Finally, following are the parameters and their values set in order to maximize performance.
\begin{itemize}
    \item Number of heads in multi-head attention layer = 5
    \item Number of convolution kernel windows = 3
    \item Number of epochs = 100
    \item Batch Size = 64
    \item Learning Rate = ${10^{-4}}$
    \item Optimizer = Adam
\end{itemize}

In experiment 2, model was compared with existing vulnerability detection tools. It is concluded from this experiment that Multi-task learning based model consistently outperforms other tools such as SmartCheck, Securify and Mythril in three dominant types of vulnerabilities namely arithmetic vulnerability, reentrancy and unknown address. F1-score for multi-task based model is 0.81, 0.74 and 0.83 respectively for each vulnerability whereas for other the range of F1-score is between 0.4 - 0.56. Thus, it is evident that multi-task learning based model is highly efficient against other non-ML tools. Next, through experiment 3, similar trend is observed that the current model outperforms various ML-based models. Again the range (for F1-score) of current model for three vulnerabilities is 0.7 - 0.88 whereas for others F1-score lies between 0.4-0.77. Lastly, ablation experiment was performed. The performance metrics took a real hit when either only detection task model was considered or recognition task model was considered. Furthermore, the need for computing resources was also compared. The conclusion for resources was that the multi-task learning model not only required fewer computing resources, it provided a great F1-score and time to train the model was also less.

Despite achieving great F1-score and consistently outperforming other tools for vulnerability detection, there are certain limitations. Firstly, the model adopts hard parameter sharing method undermining the expressiveness and generalizability of the model. Furthermore, the weight settings of the model is not determined automatically. Thus, this forms the basis of future work in multi-task learning for vulnerability detection of smart contracts. 

\subsection{Fuzzing Based analyser}
There are several tools that follow a random testing i.e a fuzzing approach. In this section we would first discuss ReGuard\cite{liu2018reguard} followed by a brief discussion of variations that can be observed in different tools using the same approach. 
ReGuard\cite{liu2018reguard} is a dynamic fuzzing based analyzer that automatically detects reentrancy bugs in Ethereum Smart Contracts. The solution provided by ReGuard solves two challenges. It is able to cover different transaction scenarios and encode reentrancy property and it is extended upon existing fuzzing engines AFL\cite{[lcamtuf.coredump.cx]} and LibFuzzer\cite{libfuzzer}. The architecture of ReGuard consists of three key modules which are \emph{Contract Transformer}, \emph{Fuzzing Engine} and \emph{Core Detector}.\\

The following is the working of ReGuard:
\begin{itemize}
    \item It takes Binary code or source code of smart contract as input.
    \item Given the code, it parses through intermediate representation which is an abstract syntax tree. For binary code contract, the control flow graph acts as IR.
    \item ReGuard performs a source-to-source transformation from IT to C++ in the \emph{Contract Transformer} phase while maintaining the same behaviour of the contract.
    \item An embedded fuzzing engine iteratively generates random bytes using runtime coverage as feedback. C++ smart contract, transactions and runtime library combined, ReGuard on execution of the contract dumps trace analysis-relevant operations such as function call and return, access to storage, branching operations and and other blockchain parameters.
    \item This trace is in turn passed on to \emph{Core detector} for reentrancy analysis and bugs are reported.
\end{itemize}
ReGuard is instantiated for Ethereum contracts. It was able to identify 7 bugs in five modified smart contracts along with avoiding false positives and negatives. Reguard is compared with Oyente. Based on the evaluation done, it is observed that while Oyente provides better overall test on smart contracts and checks for a set of bug patters, ReGuard was better at avoiding false positives and negatives in finding bugs.

There are certain variations to the base method of Reguard. These variations are listed below.
\begin{itemize}
    \item sFuzz\cite{nguyen2020sfuzz} which is a tool built on Aleth\cite{github}. This tool is extensible to Ethereum Virtual machines, oracles as well as fuzzing strategies. This tool complements Oyente. In this tool, a feedback-Guided adaptive testing is to transform generation problem into an optimization problem and use some form of feedback as an objective function in solving this problem. Methods such as crossover and mutation are used to generate new-test cases in accordance with the problem at hand.
    \item Harvey\cite{wustholz2020harvey} Harvey uses a grey-box fuzzing method which lies in middle of spectrum in terms of performance and effectiveness in discovering new-paths. One such state-of-the-art example is American Fuzzy Lop (AFL). It has analyzed more than 3 million smart contracts. It is a closed-source but is usable via a MythX\cite{mythx} CI platform and the solfuzz\cite{muellerberndt} tool.
    \item HFContractFuzzer \cite{ding2021hfcontractfuzzer} is a tool, unlike others that we over viewed, is built for smart contracts on Hyper-ledger Fabric\cite{androulaki2018hyperledger}. It uses a similar fuzzing technology along with a Go-language testing tool called go-fuzz. The smart contracts and test cases are provided by users and they are given automatic results. It also briefly mentions methods for optimization of go-fuzz for smart contracts - namely initial corpus optimization and mutation process optimization.
    \item Echidna\cite{grieco2020echidna} is an effective, easy-to-use, and fast fuzzer on Ethereum blockchain smart contracts. In addition to the in-built fuzzing sequences it allows customisation by user. Along with similar features like other competitors, it also provides an estimation of maximum gas usage which finds an application in security audits of contracts. The custom might be detected up to $63\%$ of most exploitable and severe flaws. Built on slither. Echidna inherits concepts from concept based fuzzing Echidna is designed in such a way to make continuous improvements and ensuring that current features are not degraded by this improvement.
    \item ContractFuzzer\cite{jiang2018contractfuzzer} is a comprehensive fuzzing framework made to detect 7 types of Ethereum smart contract vulnerabilities. It has lower amount of false positives and detects more types of vulnerabilities than Oyente. In comparison with Echidna, it seems that it was not able to produce any useful output even when given a four hour timeout period.
    \item AgSoLT \cite{driessen2021automated}AgSoLt works with both a random testing approach (fuzzing approach) and a guided-search approach(DynaMosa\cite{panichella2017automated} genetic algorithm). DynaMosa outperfomed fuzzer for achieving branch coverage. However, unlike other tools the fuzzer is not faster in AgSoLT despite skipping a few extra steps of selection, cross-over and mutation.
\end{itemize}

\subsection{Game Theory for Validation of Smart Contracts}
According to \cite{bigi2015validation}, the combination of game theory and formal can be used for the analysis of smart contracts. This is demonstrated in [REF NO.] by validation of DSCP, an idealised protocol for decentralised smart contract which in turn is inspired by BITHALO. This is shown by taking all the possible outcomes of the transactions. Several symbols have been used to represent the edges of the transaction such as \emph{P} denotes Pay, \emph{S} denotes sell while \emph{C}, \emph{D} denotes to confirmation or denying of satisfaction and \emph{L} denotes leaving the system without either paying or selling.\\
The game theory provided an insight on the behaviour of players in game (or protocol) by considering all possible cases and reducing them by backwards induction and considering cases only with maximum returns under the assumption of perfect rationality.\\

The results were then replicated in a formal model for automated validation based on establishing probability for DSCP by MDP (Markov's decision process) . \\ Subsequently, formal methods points out the cases in which players are unlikely to agree on the contracts. Here it takes several profile and cases into consideration for analyses for the validation of DSCP. This is done by Probabilistic CTL(Computation Tree Logic) where various values of r (a degree of honesty based on deposit and price). The maximum probability of the loss to occur is considered and a case where low probability encouraging a fraudulent behaviour is depicted. 

\subsection{Symbolic checking Tools}
In this section we discuss an overview of several tools that uses symbolic checking approach for anlyses of Smart contract. Following are tools that uses various approaches used in Symbolic checking tools.
\begin{itemize}
    \item Mythril\cite{sayeed2020smart} is an an open-source tool which takes advantage of the symbolic execution technique in order to determine the errors in code. It identifies possible reasons of the vulnerability by analyzing input transactions. Furthermore, it helps in mitigating exploitation and program vulnerability. It helps in identifying bugs and provides a highly accurate results. Its only limitation is that it is unable to extend over taints analysis over memory field and complicated patterns.
    \item Oyente\cite{sayeed2020smart} is a symbolic execution tool, to identify security loopholes in contracts that can cause possible threats. This method symbolically represents the nature of an execution path as a mathematical formula. If both formulas are valid simultaneously then a comparison is made between new and old formulas comprising of ordinary bugs. The limitation of Oyente is that it has less accuracy, provides high amount of false positives and often fails to detect critical vulnerabilities.
    \item Manticore\cite{mossberg2019manticore} is a symbolic checker that traces inputs that terminate a program, logs instruction-level implementation, and provides access to its analysis engine through Python API. The limitations of Manticore include the inability to detect contract suicide and time manipulation. Its implementation is sluggish and misses over a few vulnerabilities.
    \item Verx\cite{permenev2020verx} is a tool which can automatically prove temporal safety properties of Ethereum smart contracts. This involves a three step process which involves reduction of temporal safety verification to reachability check, conversion to identify precise symbolic states and converting those states into abstract states at the end of transaction. It finds an application as an effective tool for verifying custom functional properties of smart contracts.
    \item SmartCheck\cite{tikhomirov2018smartcheck,ferreira2020smartbugs} is a static analysis tool that is very alike to other tools in terms of approach. It builds an intermediate representation(IR), such as abstract syntax tree or three-address code in order to perform a deeper analysis compared to analyzing test. This is followed by enriching IR with additional information using control and data flow algorithms along with taint analysis, symbolic execution and abstract implementation. The vulnerability is detected running the Ir through the database of patterns. This tool needs improvement in having more precise patterns, improvement in grammar, implementation of more static analysis methods and the need for support of other languages.
    \item Zeus\cite{kalra2018zeus} utilizes symbolic model checking for analyzing safety of smart contracts. It generates XACML styled template from a smart contract. The Smart contracts along with policy specifications are translated to LLVM bitcode to enhance the contract’s behavior. Zeus performs static analysis of the furnished smart contract code to append the assert statement policy at the right spot of the program. The limitations include the lack of involving mathematical equations, no cross-function analyses, less runtime parameters.
    \item Osiris\cite{torres2018osiris} is a framework specific for detection of integer bugs in Ethereum smart contracts. This paper mainly detects three types of integer bugs which are \emph{Arithmetic bugs}, \emph{Truncation Bugs} and \emph{Signedness bugs}. Osiris works on symbolic execution and taint analysis. However, the bugs by Osiris are semantically possible, it is unlikely that such bugs are likely to be exploited in practice.
    \item Mythx\cite{mythx} includes static, dynamic, as well as symbolic execution. Mythx is integrated with truffle and Remix. The code is to be submitted after which the analysis techniques are activated and analysis reports are subsequently generated. The short comings include examination of fewer vulnerabilities.
    \item MAIAN\cite{nikolic2018finding}, unlike other tools, characterises three classes of trace vulnerabilities. MAIAN provides with high-order properties that admit symbolic analysis for detection from smart contract bytecode. This tool will be able to identify violations of \emph{safety} and \emph{liveness} properties. These traces have been analysed on smart contracts of multiple types. 
    \item FEther\cite{yang2019fether} supports hybrid symbolic execution of smart contract formal verifications. It is based on the GERM model to simulate execution patterns of solidity in Coq which guarantees consistency source code and formal models.Several mechanisms such as Ltac\cite{ltac} mechanism of Coq are automated for evaluation FEther during execution processes.
\end{itemize}

\subsection{Static Analyses tools}
In this section we discuss an overview of tools that has Static Analysis as its primary approach
\begin{itemize}
    \item EtherTrust\cite{grishchenko2018ethertrust} like static analysis tool like Zeus. However, unlike Zeus, this tool does not use Symbolic checking for its execution. It is based on Horn-clause resolution which handles the complexity of EVM bytecode in large scales. This has been tested on benchmark code snippets from literature as well as real-life contract. EtherTrust outperforms Oyente significantly, giving results in few seconds.
    \item Slither \cite{feist2019slither} is an open source static analysis framework. However, it provides fast, robust, accurate and rich information about smart contracts. It also leverages SlithIR which is an intermediate representation for practical analysis on Solidity code. Along with finding bugs, it can also be used to suggest code optimizations and increase code understanding of a smart contract.
    \item SolAnalyser\cite{akca2019solanalyser,8945725} is a tool which uses static analysis to assess location prone to vulnerabilities. This is done by execution trace analysis and code instrumentation. The source code with assertions act as correctness property check. It also supports test generator, runtime monitoring and Fault seeding tool(for analysing effectiveness of SolAnalyser with other tools) and Empirical evaluation.
    \item Rattle\cite{crytic} is a static analysis framework that takes EVM byte string and uses a flow-sensitive analysis in order to recover original control graph. This lifts CFG into an SSA/infinite register form and optimizes the SSA. Further, it removes $60\%+$ EVM instructions to make the smart contracts much more frendlier to read.
\end{itemize}

\subsection{Miscellaneous Tools}
In this section we have covered tools that do not fall under broad category or are using a slightly deviated approach from that of traditional one.
\begin{itemize}
    \item TeEther\cite{krupp2018teether} uses a complex architecture in order to identify exploits. It produces a control flow graph from EVM bytecode.This is followed by providing critical instructions, generation of path and constraints and all the possible exploits from the previous parameters. TeEther was able to identify 915 vulnerabilities while analysing 38,757 contracts from Ethereum Blockchain. A very similar could be observed in the methodology of \emph{sCompile}\cite{chang2019scompile}. However it is focused on creating a set of money-based properties based on existing vulnerabilities and identify every path which might have a potential to violate certain properties.
    \item VeriSmart\cite{so2020verismart} provides a verification algorithm which is exhaustive and takes ever behaviors of the program into account. It proposes a CEGIS-style algorithm (Counter example-guided inductive synthesis) for discovering transaction invariants during the verification process. Unlike traditional state-of-the-art bug finders, it has a rate of negligible false positives ($0.41\%$). This also depicts that the algorithm of VeriSmart overcomes the shortcomings of existing verification algorithms.
    \item Isabelle/HOL\cite{amani2018towards} splits the contracts into structure of basic blocks and program logic extracted from these into level of instruction. The proof automation tactic provides a witness of an unspecified part of the state during the post-condition of the correctness statement during the formal verification of smart contracts. Here heuristics are also defined to distinguish sites and stack unwinding from other stack manipulating instructions. Hence this tool provides a sound programming abstraction.
    \item Ethainter\cite{brent2020ethainter} is an analyzer that checks information flow along with data sanitization in order to identify composite attacks which has the potential of involving tainted information with multiple transaction to exploit severe violations. Ethainter balance of completeness and precision offers significant advantage over Securify and teEther.
\end{itemize}

\begin{table*}[]
\centering
\caption{Comparison of various framework using ML techniques. Note that LSTM-based approach is not tested for each specific vulnerability and hence is not included in this table.}
\label{tab:withML}
\resizebox{\textwidth}{!}{%
\begin{tabular}{|l|l|l|l|l|}
\hline
Attack                      & Metrics  & Graph Neural Network         & Transfer based & NLP based           \\ \hline
\multirow{2}{*}{DDos}       & F1 score &                              &                &                     \\ \cline{2-5} 
                            & Accuracy &                              & 0.96(ESCORT)   &                     \\ \hline
\multirow{2}{*}{RE}         & F1 score & 0.78(GNN), 0.86(GNN with EK) &                & 0.95(Contract Ward) \\ \cline{2-5} 
                            & Accuracy & 84.48\%                      &                &                     \\ \hline
\multirow{2}{*}{HE}         & F1 score &                              &                &                     \\ \cline{2-5} 
                            & Accuracy &                              &                &                     \\ \hline
\multirow{2}{*}{TOD}        & F1 score & 0.79                         &                & 0.96 (ContractWard) \\ \cline{2-5} 
                            & Accuracy & 83.45\%                      & 0.92(ESCORT)   &                     \\ \hline
\multirow{2}{*}{TSD}        & F1 score & 0.87(GNN with EK)            &                & 0.95 (ContractWard) \\ \cline{2-5} 
                            & Accuracy &                              &                &                     \\ \hline
\multirow{2}{*}{Arithmetic} & F1 score &                              &                &                     \\ \cline{2-5} 
                            & Accuracy &                              &                &                     \\ \hline
\end{tabular}%
}
\end{table*}

\begin{table*}[]
\caption{Comparison of various frameworks using Non-ML techniques.}
\label{tab:nonML}
\resizebox{\textwidth}{!}{%
\begin{tabular}{|l|l|l|l|l|l|}
\hline
Attack                      & Metrics  & Symbolic Checking                                                                                                                & Fuzzing Algorithm                                                                        & Static Analysis    & Sematic Analysis \\ \hline
\multirow{2}{*}{DDos}       & F1 score &                                                                                                                                  &                                                                                          &                    &                  \\ \cline{2-6} 
                            & Accuracy & 0\% (Mythril, Oyente)                                                                                                            &                                                                                          &                    & 0\%(Securify)    \\ \hline
\multirow{2}{*}{RE}         & F1 score & 0.51(Mythril), 0.45(Oyente)                                                                                                      &                                                                                          & 0.88(EtherTrust)   & 0.53(Securify)   \\ \cline{2-6} 
                            & Accuracy & 62\%(Mythril, Oyente), 100\%(Zeus), 62\%(SmartCheck)                                                                             & 100\%(sFuzz,ContractFuzzer)*                                                             &                    & 62\%(Securify)   \\ \hline
\multirow{2}{*}{HE}         & F1 score &                                                                                                                                  &                                                                                          &                    &                  \\ \cline{2-6} 
                            & Accuracy &                                                                                                                                  &                                                                                          &                    & 79\%(Securify)   \\ \hline
\multirow{2}{*}{TOD}        & F1 score &                                                                                                                                  &                                                                                          &                    &                  \\ \cline{2-6} 
                            & Accuracy & 27.1\% (Oyente), 98.93\%(Zeus)                                                                                                   &                                                                                          &                    &                  \\ \hline
\multirow{2}{*}{TSD}        & F1 score & 0.175 (Oyente), 0.08(SmartCheck)                                                                                                 &                                                                                          & 0.86 (SolAnalyser) & 0.037 (Securify) \\ \cline{2-6} 
                            & Accuracy &                                                                                                                                  & \begin{tabular}[c]{@{}l@{}}86\%(sFuzz)*, \\ 96.05\%(ContractFuzzer)*\end{tabular}        &                    &                  \\ \hline
\multirow{2}{*}{Arithmetic} & F1 score & \begin{tabular}[c]{@{}l@{}}0.12 (Oyente), 0.38 (Mythril - Underflow),\\ 0.45 (Mythril - Overflow), 0.19(SmartCheck)\end{tabular} &                                                                                          & 0.71 (SolAnalyser) & 0.76(Securify)   \\ \cline{2-6} 
                            & Accuracy & 50\% (Mythril), 55\%(Oyente)                                                                                                     & \begin{tabular}[c]{@{}l@{}}100\%(sFuzz-Overflow)*,\\ 80\%(sFuzz-Underflow)*\end{tabular} &                    &                  \\ \hline
\end{tabular}%
}
\end{table*}

\section{Conclusion}
Smart contracts enable users to form decentralized digital agreements without the need for a third party \cite{sayeed2020smart}. As mentioned earlier, the usage of smart contracts is increasing in every field including health, finance, business management etc. However, with the increasing usage, these programs are becoming more vulnerable to attacks resulting in severe loss of money.

In this paper, we provided brief about various types of attacks and vulnerabilities that are frequently observed in smart contracts. We also discussed about the frameworks and approaches for detecting such vulnerabilities. Lastly, we provided a table of comparison of these different approaches. Even though most of these frameworks provide really great performance, smart contracts are not fully immune to attacks mainly because of limitations in these frameworks.

To begin with, most of the frameworks except \cite{lutz2021escort} lack generalizability. As a result, whenever a new vulnerability appears, these approaches usually fail to capture it which can result into a huge exploitation. Next, these approaches lack a unified evaluation method. Most of these frameworks have been evaluated on their personal database collected from solidity or etherscan. All these datasets are different and hence directly comparing these results might not give us an overall view of which framework is better. Finally, almost all of these frameworks are able to detect a subset of vulnerabilities. Even though this subset contains critical vulnerabilities, there are some vulnerabilities that these frameworks miss out.

However after analysing 39 frameworks, we were able to summarize our findings using frequently occurring  metrics accuracy and f1-score. Table \ref{tab:withML} summarises the findings for tools using machine learning approaches. This tools were covered in \emph{A} to \emph{M} in section \ref{section3}. While Table \ref{tab:nonML} summarises the findings for the tools using non-machine learning approaches which were covered in \emph{N} to \emph{R} of section \ref{section3}. 

The table has various empty entries indicating either a lack of availability of statistics or the inability of tool to cover a particular attack. Along with this many tools were not included here because of lack of unified metrics to establish context amongst them. This could establishes a future work for an in-depth experimentation for a thorough comparison.

Many approaches such as analysis using game theory are merely theoretical which holds a potential application in analysing smart contracts. It is also worth mentioning that the tools taken into consideration form comparison often finds various specific applications and hence the selection of such tools should be based on various factors other than the suggested metrics.

\bibliographystyle{plain}
\bibliography{Citations.bib}

\vspace{12pt}
\end{document}